\documentclass{trbunofficial}

\usepackage{indentfirst}
\usepackage{tikz}
\usepackage{amsmath}
\usepackage{bm}
\usepackage{enumitem}
\usepackage{algorithm}
\usepackage{algpseudocode}
\usepackage{hyperref}
\usepackage{booktabs, graphicx}
\usepackage{tabularx}
\usetikzlibrary{fit,positioning,backgrounds,arrows.meta}

\begin{document}


\title{Transit Destination Inference from Tap-In-Only Bus Smart-Card Data: A Hierarchical Bayesian Approach}


\TRBauthor{Gefei Zhao}
{Department of Computer Science}
{University of Wisconsin--Madison, Madison, WI, 53706}
{}
{gzhao46@wisc.edu}

\TRBauthor*{Jiahe Ling, M.S.}
{Department of Industrial Engineering and Operations Research}
{Columbia University, New York, NY, 10027}
{https://orcid.org/0000-0001-9106-1375}
{jl7264@columbia.edu}

\TRBauthor{Yuelong Su, Ph.D.}
{State Key Laboratory of Intelligent GreenVehicle and Mobility}
{Tsinghua University, Beijing, China, 100084}
{https://orcid.org/0009-0006-9645-5158}
{suyuelong@btifii.org.cn}

\AuthorHeaders{\textit{Zhao, Ling, and Su}}


\maketitle 

\section{Abstract}
\hfill\break%
\noindent\textbf{Objectives:}~Entry-only fare systems record boardings but
not alightings, preventing direct construction of origin--destination (OD)
matrices. This study develops a Hierarchical Bayesian Latent-Destination
(HBLD) model that combines historical station--hour boarding patterns with
card history, treats trip-chain destinations as uncertain evidence, and
propagates destination uncertainty into OD flows.

\hfill\break%
\noindent\textbf{Methods:}~This study analyzes 838{,}305 bus tap-ins from Changzhou
in May 2025, linked to network and weather data. Given the boarding stop,
HBLD estimates probabilities over downstream and through-terminal
opposite-direction stops, while trip-chain estimates enter as noisy
evidence with reliability $\pi$. It combines network, temporal, and weather
effects, negative-binomial-smoothed historical boarding and inferred
alighting patterns, and Bayesian personalization that falls back to the
shared model without card history. The model was fitted by stochastic
variational inference and evaluated out of time on the final week using
component-removal tests, day-clustered bootstrapping, and sensitivity
analysis across $\pi$, and was scored against trip-chain outputs because
true alightings are unavailable.

\hfill\break%
\noindent\textbf{Findings:}~HBLD outperformed the strongest baseline by
improving the destination distribution rather than the most likely stop.
Observed boarding patterns improved prediction across reliability settings,
particularly without card history, whereas inferred alighting patterns
helped only when trip-chain evidence was strongly trusted. Destination
rankings remained stable at high reliability, but OD flows were more
sensitive, and HBLD captured through-terminal and bus-assisted road-crossing
behavior while extending OD estimation to trips unresolved by deterministic
chaining.

\hfill\break%
\noindent\textbf{Novelty:}~HBLD integrates observed boarding aggregates
with individual trip inference while treating trip chaining as noisy
evidence, distinguishing robust observed information from chain-derived
aggregate patterns.

\hfill\break%
\noindent\textbf{Practical Applications:}~Operators can estimate a
destination distribution at tap-in and aggregate these predictions into
uncertainty-aware OD matrices. The outputs support operations, planning,
scheduling, resource allocation, terminal management, and pedestrian-access
improvements while including trips that deterministic chaining cannot
resolve.

\newpage

\section{Introduction}\label{sec:intro}

\noindent Knowing where passengers travel, rather than merely where they board, is fundamental to effective public transport planning and operations. Complete origin--destination (OD) flows reveal travel patterns and guide network design, scheduling, resource allocation, and service evaluation \citep{tang_incorporating_2020}. Stop-level alighting estimates also enable detailed trip, route, and mobility-pattern analyses that can inform service improvements \citep{cerqueira_moving_2024}. Yet many automatic fare collection (AFC) systems, particularly flat-fare entry-only systems, record boarding locations but not alighting locations \citep{egu_how_2020,cui_alighting_2021}. This limitation occurs across diverse transit contexts, including systems in London, Santiago, and Chinese cities \citep{yap_improving_2018,cui_alighting_2021}, leaving each recorded trip without an observed destination. Missing alighting records in entry-only AFC systems have therefore motivated various destination-inference methods.

Trip chaining remains the dominant approach for inferring missing alighting stops. It assumes that each trip ends near the passenger’s next boarding stop and that the final daily trip ends near the day’s first boarding stop \citep{barry_origin_2002,trepanier_individual_2007}. Subsequent studies refined this approach using operational and historical information, while probabilistic and machine-learning methods extended inference to unlinked trips \citep{nassir_transit_2011,he_estimating_2015,cheng_probabilistic_2021,cerqueira_moving_2024}. Aggregate methods have separately reconstructed system-level OD flows from passenger counts \citep{mohammed_origin-destination_2023}. Despite this progress, two gaps remain. First, existing individual destination models do not incorporate aggregate OD information into passenger-level predictions. Second, when true alighting stops are unavailable, supervised models often treat trip-chain estimates as true training labels \citep{tang_incorporating_2020,cui_alighting_2021}, allowing chaining errors to propagate into model predictions \citep{assemi_improving_2020}. These gaps motivate a framework that integrates aggregate and individual information while treating trip-chain destinations as uncertain rather than observed.

To address these gaps, this study develops a Bayesian framework that integrates aggregate demand information with individual trip records and treats trip-chain destinations as uncertain evidence for latent alighting stops. The framework estimates destination probabilities for both chainable and unlinked trips, allowing uncertainty to propagate into aggregate OD flows rather than treating inferred destinations as ground truth. The paper is structured as follows. The next section reviews the relevant literature, followed by the data and methods, including trip chaining and the Bayesian model. The subsequent section presents and discusses the results, and the final section summarizes the main findings, implications, limitations, and directions for future research.

\section{Literature Review}\label{sec:literature}
\noindent Transit OD estimation depends on the AFC system design. Entry--exit systems observe boarding and alighting, allowing OD pairs to be constructed after cleaning, whereas entry-only systems require destinations to be inferred from boarding records \citep{hussain_transit_2021}. This missing-destination problem has motivated several families of alighting-stop and OD estimation methods.

The dominant entry-only approach is trip chaining. Introduced by \citet{barry_origin_2002}, it assumes that passengers alight near their next boarding location and end their final daily trip near their first origin. For buses, the inferred destination is generally the downstream stop closest to the next boarding. Later studies formalized feasible downstream stops \citep{trepanier_individual_2007}, matched transactions to vehicle trips using schedules and automated data \citep{nassir_transit_2011}, combined AFC with Automatic Vehicle Location and survey validation \citep{wang_bus_2011}, tested walking and transfer thresholds \citep{alsger_use_2015}, and introduced fare-zone and duplicate-tap validation \citep{nunes_passenger_2016}. These refinements improve reliability, but trip chaining remains assumption-dependent and weak for unlinked or irregular trips.

Probabilistic and pattern-based methods reduce this dependence by comparing feasible destinations or exploiting historical regularity. \citet{kumar_robust_2018} ranked candidate trajectories using probabilities for vehicle-location error, schedule deviation, and route choice, increasing inference coverage from 70\% to 85\%. \citet{he_estimating_2015} applied spatial and temporal kernel densities to unlinked trips, while \citet{cheng_probabilistic_2021} represented time, origin, and destination through latent travel-behaviour topics. Related clustering methods combine K-means with Gaussian mixtures \citep{lee_travel_2022} or trip chaining with DBSCAN and frequent-pattern mining \citep{cerqueira_moving_2024}. These methods improve coverage for regular passengers but are less effective when travel histories are sparse or unstable.

Machine-learning methods learn nonlinear relationships between destinations and passenger, trip, network, and contextual attributes. With observed entry--exit labels, \citet{jung_deeplearning_2017} trained a neural network using trip, network, and land-use features, while \citet{assemi_improving_2020} used neural networks to correct trip-chaining errors, increasing exact-stop accuracy from 72.2\% to 79.5\%. Without observed exits, \citet{tang_incorporating_2020} trained gradient-boosted trees on trip-chain-generated labels, travel history, and weather, and \citet{yan_alighting_2019} combined rule-based inference with ensemble classification. \citet{hamedmoghadam_automated_2021} further integrated candidate generation, random-forest selection, and data-driven transfer identification. These models increase flexibility, but those trained on pseudo-labels can reproduce trip-chaining errors, while models trained on entry--exit data may not transfer directly to entry-only systems.

Whereas these methods infer individual destinations, classical aggregate methods reconstruct OD matrices from passenger counts. Iterative proportional fitting adjusts a seed matrix to match observed margins, while maximum entropy selects a feasible matrix with minimal additional information \citep{mohammed_origin-destination_2023}. Both are efficient but sensitive to the seed matrix and require alighting, link-flow, or other destination-related totals.

Bayesian approaches extend aggregate reconstruction while also supporting individual sequential modelling. For general networks, \citet{li_bayesian_2005} used a Poisson model and Expectation--Maximization (EM) algorithm to update prior OD flows from link counts. At the transit-route level, \citet{li_markov_2009} imposed a Markov structure on OD flows and combined prior information with boarding and alighting counts, while \citet{hazelton_statistical_2010} used Markov chain Monte Carlo to sample feasible OD matrices. At the individual level, \citet{zhao_individual_2018} developed a Bayesian N-gram model for next-trip time, origin, and destination, and \citet{mo_individual_2022} used an input--output hidden Markov model to represent latent activities and jointly predict next-trip time and origin. For larger systems, \citet{blume_bayesian_2022} estimated posterior OD proportions from network-wide inflow and outflow counts using Hamiltonian Monte Carlo. Finally, \citet{chen_bayesian_2025} modelled time-varying bus OD matrices through multinomial alighting probabilities, low-rank temporal factors, Gaussian-process priors, and Markov chain Monte Carlo. These models demonstrate the Bayesian advantages of prior incorporation, latent-state representation, information sharing, and uncertainty quantification, but use link counts, boarding and alighting totals, or complete historical OD records.

Few studies jointly use aggregate passenger counts and individual trip attributes to infer destinations, and the machine-learning approaches that do so offer limited interpretability. Such integration is particularly valuable for entry-only AFC, where destinations are often inferred by trip chaining and then used as labels to model unresolved or all trips \citep{tang_incorporating_2020,yan_alighting_2019}. Because these labels are approximate, treating them as ground truth can propagate trip-chaining errors \citep{assemi_improving_2020}. This motivates an interpretable joint model that uses aggregate counts to inform individual predictions while treating trip-chain destinations as uncertain observations and quantifying alighting uncertainty.

\section{Methods}\label{sec:methods}
\begin{figure}[htbp]
    \centering
    \includegraphics[width=0.8\textwidth]{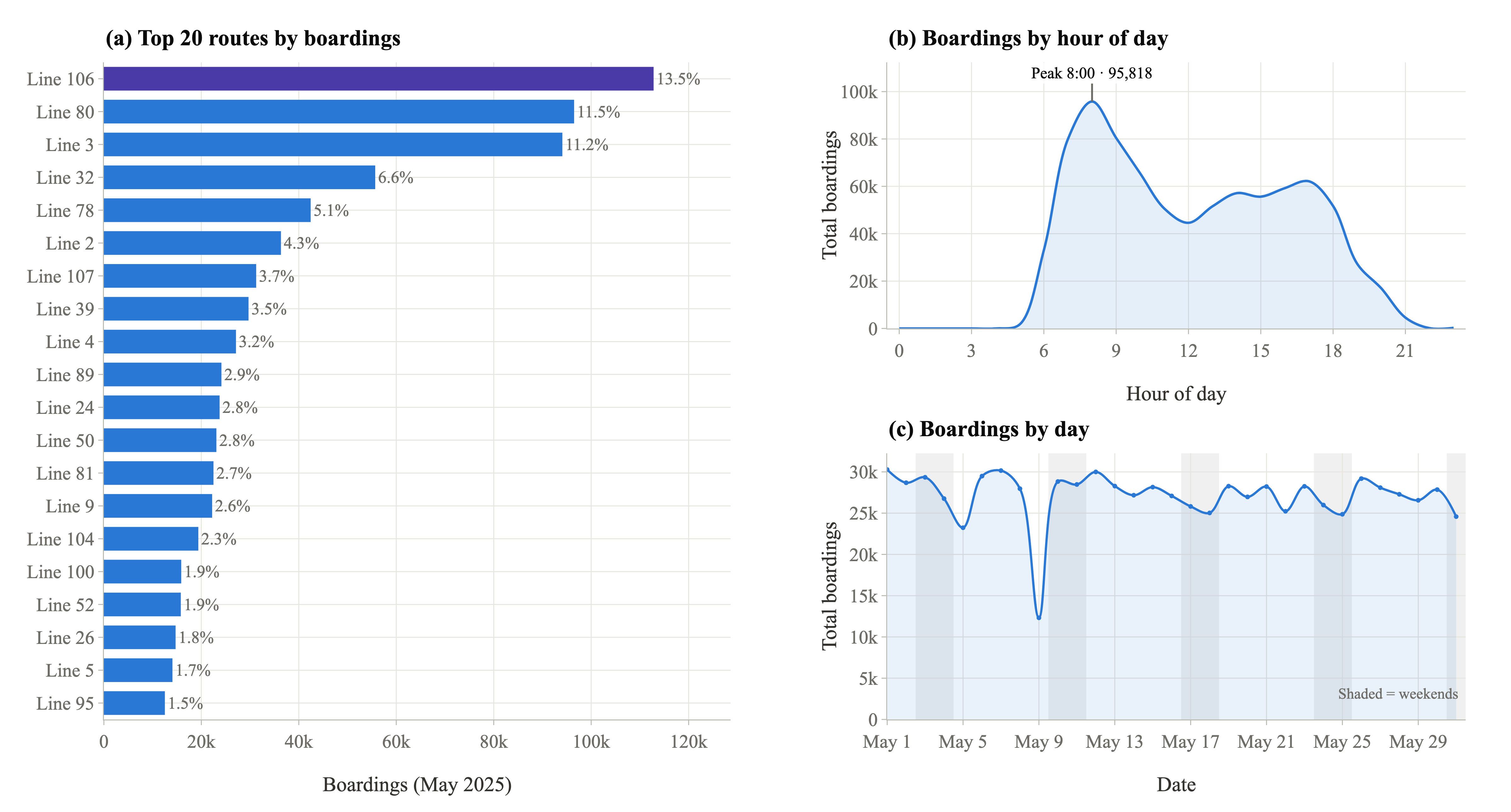}
    \caption{Transit Boarding Demand Patterns.}
    \label{fig:demand_overview}
\end{figure}

\begin{figure}[htbp]
    \centering
    \includegraphics[width=1\textwidth]{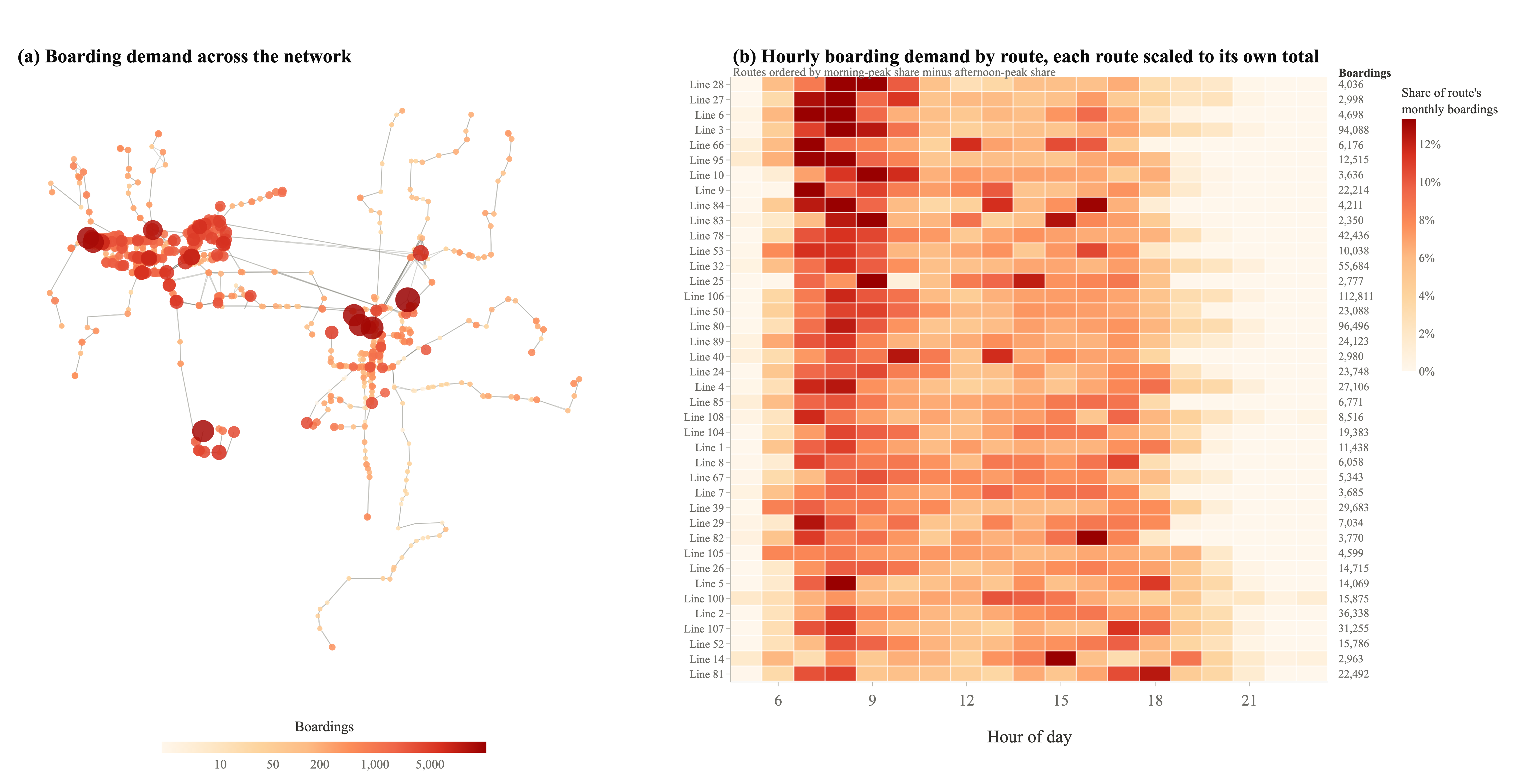}
    \caption{Spatial-temporal Boarding Demand.}
    \label{fig:demand_space_time}
\end{figure}

\noindent This section discusses the data, trip-chaining, and Hierarchical Bayesian Latent-Destination (HBLD) model. Figures~\ref{fig:demand_overview} and~\ref{fig:demand_space_time} show the demand patterns, and Figure~\ref{fig:hbld_graph} explains the model structure.

\subsection{Data}

\noindent The study uses bus smart-card transaction data from Changzhou, China, covering May 1–31, 2025. Changzhou operates a flat-fare, entry-only system: passengers tap their cards when boarding but not when alighting. Each transaction therefore records the card identifier, route, direction, boarding stop and its sequence position, and boarding time, but not the alighting stop.

Data cleaning proceeded as follows. Records with all-zero card identifiers were removed, followed by records whose stop keys were absent from the cleaned stop list. Cards with only one remaining transaction were excluded because their trips could not be chained. Duplicate transactions by the same card at the same time and repeated taps within two minutes were then resolved. Because these steps left some cards with only one transaction, such cards were removed again. Finally, all retained records were required to have valid route, direction, stop position, time, and stop-coordinate fields. Of the 883,897 raw transactions, 838,305 (94.8\%) were retained, representing 84,502 cards. The median card recorded four boardings during the study month.

The cleaned dataset covers 43 routes and 962 stations. A directional route contains an average of 20.8 stops and a maximum of 41, thereby bounding the candidate-stop sets defined below. Demand is highly concentrated across routes: the three busiest routes account for 36.2\% of all boardings, and the ten busiest account for 65.6\%. Demand also varies substantially by time of day. The system records an average of 27,042 boardings per day, with service operating approximately from 05:00 to 23:00. The morning peak is pronounced, reaching 95,818 boardings during the 08:00 hour; the 07:00–09:00 period accounts for 30.6\% of total demand. The afternoon peak is lower and more dispersed, with 20.6\% of boardings occurring between 16:00 and 18:00.

Hourly weather observations for May 2025 were obtained from Meteostat station 58343 in Changzhou (31.7667\textdegree N, 119.9500\textdegree E). Following Tang et al. (2020), who found temperature and weather type to be more predictive of alighting stops than precipitation, humidity, visibility, or wind speed, only these two variables were retained. Weather type was classified into four categories: clear or fair, cloudy or overcast, light rain, and moderate-to-heavy rain. All 838,305 transactions were successfully matched to hourly weather observations, with no missing matches.


\subsection{Trip Chaining}

\noindent  In tap-in-only AFC systems, boarding records do not reveal alighting stops, so inference requires behavioral assumptions. Prior studies have used trip-chain-inferred alighting stops as training labels for supervised destination models \citep{tang_incorporating_2020,cui_alighting_2021}. Trip chaining assumes that each trip ends near the next observed boarding stop and that the final trip of the day ends near the day’s first boarding stop. However, deterministic trip chaining returns a single destination and cannot represent uncertainty arising from nearby alternatives, incomplete chains, transfers, loop routes, route variants, or intermediate non-transit travel. This study therefore uses trip chaining as a behavioral signal, rather than ground truth, within a Bayesian framework that treats the alighting stop as latent and estimates posterior probabilities over candidate downstream stops.

Alighting locations are inferred from each card's chronologically ordered boardings (\autoref{alg:tripchaining}). For a boarding $b$ at station $s(b)$ and position $k(b)$ on route $(\ell(b),d(b))$, $\mathcal{A}(b)$ contains same-direction stops strictly downstream of $k(b)$, whereas $\mathcal{B}(b)$ contains the full opposite-direction sequence $(\ell(b),1{-}d(b))$ to accommodate onward opposite-direction travel, including through-terminal riding and bus-assisted road crossing; both sets exclude the boarding station. Given a target station $s^{*}$, the candidate with the smallest haversine distance to $s^{*}$ is accepted if it lies within $M=400$\,m. The algorithm searches $\mathcal{A}(b)$ first and searches $\mathcal{B}(b)$ only if no match is found, yielding \textit{same-direction chaining} and \textit{opposite-direction chaining}, respectively. Building on the trip-chaining algorithm of \citet{trepanier_individual_2007}, inference proceeds sequentially under three rules: (i)~for every boarding except the last of the service day, $s^{*}$ is the immediately following boarding station, on any line; (ii)~for the day's last boarding, $s^{*}$ is the first boarding station of that day, imposing daily-tour closure, with the first boarding station on the next calendar day used if no match is found; and (iii)~remaining boardings are imputed only when the same card has spatially chained trips sharing the same line, direction, boarding station, and weekday, with boarding times within one hour of the same time of day, and the modal destination among these eligible trips is then assigned (\textit{recurrence-based imputation}).

Of 838{,}305 boardings made with 84{,}502 cards, 80.4\% were successfully chained, comparable to the rate reported by \citet{cheng_probabilistic_2021}: 71.0\% by \textit{same-direction chaining}, 7.2\% by \textit{opposite-direction chaining}, and 2.3\% by \textit{recurrence-based imputation}. By rule, 47.9\% of all boardings were resolved under rule (i), 30.3\% under rule (ii),
and 2.3\% under rule (iii); the remaining 19.6\% were left unchained rather than imputed from weaker evidence. The median matching distance is 45\,m for both spatial chaining categories, indicating comparable spatial precision. 

Prior alighting-inference studies account for opposite-direction travel \citep{cerqueira_inference_2022,nassir_transit_2011}. Operations staff also report that passengers near terminals routinely board arriving inbound vehicles and remain aboard through the turnaround. Three results support through-terminal riding. First, final stops have no downstream same-direction candidates, yet 83.2\% of 9{,}609 final-stop boardings (1.1\% of the sample) match an opposite-direction stop, with a median matching distance of 36\,m. Second, among spatially chained trips, the opposite-direction share rises from 8.7\% in the first half of the route to 62.8\% in the final position decile (two-proportion $z=209.7$, $p<10^{-3}$). 
Third, terminal-proximal opposite-direction chains (boarding position $\ge 0.7$ of route length) travel a median of eight stops beyond the terminal, or 41\% of the opposite-direction route. Repeat cards account for 55\% of these trips, indicating purposeful and habitual behavior. 

Moreover, 47{,}460 opposite-direction chains occur mid-route and appear to represent bus-assisted road crossing. Bus operations staff suggest elderly riders using the bus to avoid crossing roads on foot. Among these trips, 72.5\% end directly opposite the boarding stop and 19.4\% end opposite an upstream stop. Their median origin--destination separation is only 70m, compared with 3{,}379,m for terminal-proximal chains, and repeat cards account for 78\% of the trips. This concentration at opposite-side stops, together with repeated use by the same cards, indicates systematic behavior rather than matching error.

\begin{algorithm}[t]
\caption{Rule-based Trip Chaining}
\label{alg:tripchaining}
\begin{algorithmic}[1]
\Require Boardings $T_c=\{b_1,\dots,b_n\}$ of each card $c$, sorted by time;
         candidate sets $\mathcal{A}(b)$, $\mathcal{B}(b)$; distance buffer $M$
\Function{Match}{$b$, $s^{*}$}
  \ForAll{$(\mathcal{C},\text{method}) \in
          \big[(\mathcal{A}(b),\textit{same-direction}),\,
               (\mathcal{B}(b),\textit{opposite-direction})\big]$}
     \State $x^{\ast} \gets \arg\min_{x\in\mathcal{C}} \delta\big(s(x),s^{*}\big)$
     \If{$\mathcal{C}\neq\emptyset$ \textbf{and} $\delta\big(s(x^{\ast}),s^{*}\big)\le M$}
        \State \Return $(x^{\ast},\text{method})$
     \EndIf
  \EndFor
  \State \Return $\varnothing$
\EndFunction
\For{each card $c$ \textbf{and} each boarding $b_i \in T_c$}
   \If{$b_i$ is \textbf{not} the last boarding of its service day}
      \State $r \gets \Call{Match}{b_i,\, s(b_{i+1})}$
   \Else
      \State $r \gets \varnothing$;\quad
             $f \gets$ first boarding of the same day
      \If{$f \neq b_i$}
         \State $r \gets \Call{Match}{b_i,\, s(f)}$
      \EndIf
      \If{$r = \varnothing$}
         \State $r \gets \Call{Match}{b_i,\, s(\text{first boarding of next calendar day})}$ 
      \EndIf
   \EndIf
\EndFor
\For{each boarding still unchained}
   \State assign modal destination of $c$'s chained trips sharing
          $(\ell,\, d,\, s,\, \mathrm{weekday})$ with circular hour distance
          $\le 1$, if any exist
\EndFor
\end{algorithmic}
\end{algorithm}



\subsection{Hierarchical Bayesian Latent-Destination (HBLD) Model}

\noindent The HBLD model is designed around three empirical requirements
and the uncertain status of trip-chain destinations. First, observed destinations from the same boarding stop are often
distributed across multiple stops, motivating estimation of a full
destination distribution rather than a single alighting prediction.
Second, rolling out-of-time analysis showed that destination patterns are
better captured by hour of day than by same-weekday history, while boarding
counts vary by day type and exhibit substantial overdispersion. We therefore
use hour-specific destination effects and an Negative Binomial (NB) demand model distinguishing
workdays from weekends and holidays. 
Third, prior same-origin card history is substantially more predictive than
population patterns. We therefore adds a partially pooled card
layer that personalizes repeated trips while reverting to the population
distribution for cold-start cards.
Trip-chain outputs are observed proxy labels rather than ground-truth
alightings, so treating them as exact outcomes can propagate chaining
errors into model estimation \citep{assemi_improving_2020}. The HBLD model
accordingly represents the true destination as latent, combines
NB-smoothed aggregate information with population and card-level
behavior, and incorporates the chain output through an explicit
noisy-evidence likelihood.

\usetikzlibrary{arrows.meta,shapes.geometric}

\begin{figure}[htbp]
\centering
\resizebox{0.9\linewidth}{!}{%
\begin{tikzpicture}[
    >=Latex,
    font=\small,
    observed/.style={
        draw,
        rounded corners,
        fill=gray!20,
        align=center,
        minimum height=9mm,
        text width=25mm
    },
    deterministic/.style={
        draw,
        rounded corners,
        align=center,
        minimum height=9mm,
        text width=25mm
    },
    parameter/.style={
        draw,
        ellipse,
        align=center,
        minimum height=8mm,
        text width=21mm
    },
    latent/.style={
        draw,
        circle,
        align=center,
        minimum size=11mm
    },
    output/.style={
        draw,
        double,
        rounded corners,
        align=center,
        minimum height=9mm,
        text width=23mm
    },
    arrow/.style={
        ->,
        line width=0.7pt
    }
]

\node[observed] (countsB) at (-4.0,4.6)
    {Historical boardings\\$N_{s,h,t<t_n}$};

\node[deterministic] (nbB) at (-0.5,4.6)
    {NB boarding\\hierarchy};

\node[deterministic] (attractorB) at (2.9,4.6)
    {Boarding attractor\\$\bar B^{(<t_n)}_{s,h,\tau}$};

\node[deterministic] (countsA) at (-4.0,2.7)
    {Historical alighting\\evidence $C^{(<t_n)}_{s,h,\tau}$};

\node[deterministic] (nbA) at (-0.5,2.7)
    {NB alighting\\hierarchy};

\node[deterministic] (attractorA) at (2.9,2.7)
    {Alighting attractor\\$\bar A^{(<t_n)}_{s,h,\tau}$};

\node[observed] (context) at (-4.0,0)
    {Trip context\\$o_n,h_n,\tau_n$\\$w_n,T_n$};

\node[observed] (network) at (-4.0,-2.8)
    {Candidate features\\$\mathcal C(o_n)$\\$D(o_n,j)$\\$r(o_n,j)$};

\node[deterministic] (population) at (-0.5,0)
    {Population choice\\$p_n(j)$};

\node[deterministic, text width=43mm] (personalized) at (4,0)
    {Card-personalized choice\\
     $q_n(j)\!=\!\Pr(y_n\!=\!j\mid c(n))$};

\node[latent] (destination) at (7.6,0)
    {$y_n$};

\node[observed] (chain) at (10.2,0)
    {Chain output, if available\\$z_n$};

\node[parameter] (theta) at (-0.5,-2.8)
    {Choice parameters\\$\bm{\theta}$};

\node[deterministic] (card) at (3,-2.8)
    {Earlier card history\\
     $n^{(<t_n)}_{c,o_n,j},\tau_{\!c}$};

\node[deterministic] (posterior) at (6.5,-2.8)
    {Posterior destination\\$\tilde p_n(j)$};

\node[output] (od) at (10.2,-2.8)
    {Conditional posterior\\OD flows $Q_{odt}$};

\node[parameter] (pi) at (10.2,2.6)
    {Chain reliability\\$\pi$};

\draw[arrow] (countsB) -- (nbB);
\draw[arrow] (nbB) -- (attractorB);
\draw[arrow, bend left=15]
    (countsB.north east) to (attractorB.north west);

\draw[arrow,rounded corners=2pt]
    (attractorB.east)
    -- ++(8mm,0)
    |- ([yshift=5mm]population.north)
    -- (population.north);

\draw[arrow, dashed, rounded corners=3pt]
    (chain.east) -- ++(7mm,0)
    -- ([xshift=7mm]chain.east |- 0,6)
    -- ([xshift=-6mm]countsA.west |- 0,6)
    -- ([xshift=-6mm]countsA.west)
    -- (countsA.west);
\draw[arrow] (countsA) -- (nbA);
\draw[arrow] (nbA) -- (attractorA);
\draw[arrow, bend right=15]
    (countsA.south east) to (attractorA.south west);
\draw[arrow]
    (attractorA.south)
    -- ([xshift=3mm]population.north);

\draw[arrow] (context.east) -- (population.west);
\draw[arrow] (network.east) -- (population.west);
\draw[arrow] (theta) -- (population);

\draw[arrow] (population) -- (personalized);
\draw[arrow] (card) -- (personalized);
\draw[arrow, dashed]
    (posterior.west) -- (card.east);
\draw[arrow] (personalized) -- (destination);
\draw[arrow, dashed, rounded corners=3pt]
    ([xshift=7mm]population.north)
    -- ([xshift=7mm]population.north |- 0,1.1)
    -- (destination.north |- 0,1.1)
    -- (destination.north);

\draw[arrow] (destination) -- (chain);
\draw[arrow] (pi) -- (chain);
\draw[arrow] (personalized.south) -- (posterior.north west);
\draw[arrow] (chain.south) -- (posterior.north east);
\draw[arrow] (posterior) -- (od);

\end{tikzpicture}%
}

\caption{HBLD estimation and OD inference. Shaded: observed data;
rectangles: deterministic or empirical-Bayes quantities; ellipses:
parameters ($\bm\theta$ estimated, $\pi$ fixed on a sensitivity grid);
circle: latent destination; double border: OD output. Dashed arrows mark
across-day flows and the estimation-time likelihood.}
\label{fig:hbld_graph}
\end{figure}
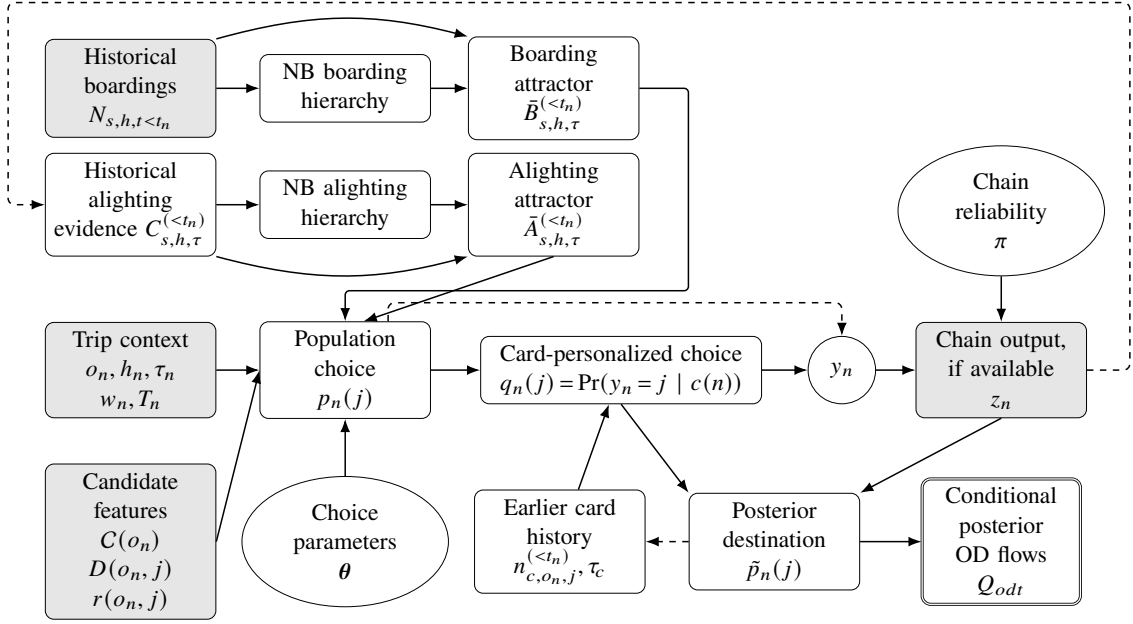

\subsubsection{Basic Notations}
\noindent Boarding $n$ occurs at directional stop $o_n=(\ell,d,k)$ (line $\ell$,
direction $d\in\{0,1\}$, sequence position $k$), station $s(o_n)$, where
$s(\cdot)$ maps any directional stop to its station, on service day
$t_n$, in hour-bin $h_n\in\{$early, AM peak, midday, PM, evening$\}$
(breakpoints taken from the observed demand profile), day type
$\tau_n\in\{\text{workday},\allowbreak\text{weekend/holiday}\}$, with
weather class $w_n\in\{$sunny/fair, cloudy/overcast, light rain,
rain/heavy rain$\}$ and temperature $T_n$ in $^\circ$C. The latent alighting stop is
$y_n\in\mathcal{C}(o_n)=\mathcal{A}(o_n)\cup\mathcal{B}(o_n)$, where $\mathcal{A}(o_n) = \{(\ell,d,j): j>k\}$ which means same direction downstream stops, $\mathcal{B}(o_n) = \{(\ell,1{-}d,j): 1\le j\le L_{\ell,1-d}\}$ which means opposite direction stops, and $L_{\ell,d}$ denotes the number of stops of line-direction
$(\ell,d)$. All stops sharing the boarding station $s(o_n)$ are excluded from both candidate sets. Each candidate carries two distinct distances:
\begin{enumerate}[
    leftmargin=1.6em,
    labelsep=0.4em,
    topsep=2pt,
    itemsep=1pt,
    parsep=0pt
]
\item In-vehicle distance $D(o_n,j)$ is the route distance travelled from
$o_n$ to candidate $j$. Let
$m(\ell,d,j)=\sum_{i<j}\delta(\text{stop}_i,\text{stop}_{i+1})$ denote
cumulative chainage, where $\delta(\cdot,\cdot)$ is the haversine distance
between consecutive stops. Define $\delta_{\text{turn}}$ as the distance from the terminal of direction $d$ to the first stop of direction $1-d$, then
\[
D(o_n,j)=
\begin{cases}
m(\ell,d,j)-m(\ell,d,k), & j\in\mathcal A,\\[2pt]
m(\ell,d,L_{\ell,d})-m(\ell,d,k)
+\delta_{\text{turn}}+m(\ell,1-d,j), & j\in\mathcal B,
\end{cases}
\]
\item Spatial displacement $r(o_n,j)$: straight-line distance
between boarding and candidate stations.
\end{enumerate}
The distinction is essential because the two can diverge sharply. Mid-route
opposite-direction trips in the data, predominantly riders using the bus
to cross the road, have a median displacement of only $\sim$70~meters while
their in-vehicle distance spans the ride to the terminal and back. Riding
cost and destination proximity are therefore separate covariates: $D$
captures in-vehicle disutility for ordinary destination-seeking travel,
while a short-range decay in $r$, active only for opposite-direction
candidates, captures the walk-saving utility of alighting close to where
one boarded. This captures both mirror-stop alightings and the 19.4\% of crossing
trips in opposite upstream stop.

\subsubsection{Generative Model}
\noindent \autoref{fig:hbld_graph} summarizes the dependency structure: the two
NB-smoothed attractors and the trip-level covariates feed the population
choice model, the card layer personalizes it, the chain output enters
only as noisy evidence on the latent destination, and the retrospective
posterior aggregates into the OD flows.
For the aggregate demand, boardings $N_{s,h,t}$ at station $s$, hour-bin $h$, day $t$ follow
\begin{equation}
N_{s,h,t}\sim\mathrm{NB}\!\left(\mu_{s,h,t},\phi_s\right),\qquad
\log\mu_{s,h,t}=\beta_0+u_s+\rho_h+\zeta_{\tau(t)}+\xi_{s,h}
+\varepsilon_t,
\label{eq:nb}
\end{equation}
with $\beta_0$ the baseline log-intensity, station effects
$u_s\sim\mathcal N(0,\sigma_u^2)$ capturing persistent volume differences
across stations, hour-bin main effects $\rho_h$ giving the system-wide
daily profile, a day-type shift $\zeta_{\tau(t)}$ where
$\tau(t)\in\{\text{workday},\text{weekend/holiday}\}$ denotes the day
type of service day $t$ ($\tau_n=\tau(t_n)$), station--hour
interactions $\xi_{s,h}\sim\mathcal N(0,\sigma_\xi^2)$ allowing
station-specific temporal profiles, a day-level shock
$\varepsilon_t\sim\mathcal N(0,\sigma_\varepsilon^2)$ absorbing
system-wide variation, and hierarchical station-specific dispersion
$\log\phi_s\sim\mathcal N(\log\phi_0,\sigma_\phi^2)$ under the
parameterization
$\mathrm{Var}(N_{s,h,t})=\mu_{s,h,t}+\mu_{s,h,t}^2/\phi_s$, so larger
$\phi_s$ implies weaker overdispersion at station $s$. Based on data, priors are defined as
$\beta_0\sim\mathcal N(0,3^2)$; $\rho_h,\zeta_\tau\sim\mathcal N(0,1^2)$;
$\sigma_u\sim\mathrm{Half}\text{-}\mathcal N(0,1.5)$,
$\sigma_\xi\sim\mathrm{Half}\text{-}\mathcal N(0,0.5)$,
$\sigma_\varepsilon\sim\mathrm{Half}\text{-}\mathcal N(0,0.2)$;
$\log\phi_0\sim\mathcal N(\log 4,1^2)$,
$\sigma_\phi\sim\mathrm{Half}\text{-}\mathcal N(0,0.7)$.
Fitted to alighting evidence, yields with
\autoref{eq:nb} the two aggregate attractors, defined as as-of-day
Dirichlet posterior-mean shares:
\begin{equation}
\bar A^{(<t)}_{s,h,\tau}
=\frac{C^{(<t)}_{s,h,\tau}+\alpha_A\,m^{\mathrm{NB},A}_{s,h,\tau}}
{\sum_{s'}C^{(<t)}_{s',h,\tau}+\alpha_A},\qquad
\bar B^{(<t)}_{s,h,\tau}
=\frac{N^{(<t)}_{s,h,\tau}+\alpha_B\,m^{\mathrm{NB},B}_{s,h,\tau}}
{\sum_{s'}N^{(<t)}_{s',h,\tau}+\alpha_B},
\label{eq:attractors}
\end{equation}
where $N^{(<t)}_{s,h,\tau}=\sum_{t'<t}N_{s,h,t'}$ are cumulative observed
boardings, $C^{(<t)}_{s,h,\tau}$ is the analogous cumulative alighting
evidence (initialized from chain outputs and later refreshed with
posterior expected counts), $m^{\mathrm{NB},A}$ and $m^{\mathrm{NB},B}$
are the corresponding NB posterior-mean intensities normalized within
each (hour-bin, day-type) cell so that $\sum_s m_{s,h,\tau}=1$, and the
concentrations $\alpha_A,\alpha_B>0$ are selected by held-out predictive
likelihood on the nested training split; each acts as an effective prior
sample size, so cells with less accumulated evidence than $\alpha$ are
dominated by the NB-smoothed profile. Both attractors use only days
$<t$, so they are safe at prediction time, and sparse or early-day cells
shrink toward the NB-smoothed profile rather than noisy raw means.

Given covariates $\bm{x}_n=(o_n,h_n,\tau_n,w_n,T_n)$ and system history
$\mathcal H_{<t_n}$ (all quantities computable from service days before
$t_n$: cumulative evidence, NB prior means, and card soft counts), the
latent destination follows a multinomial logit
over the candidate set,
\begin{equation}
p_n(j)\;\equiv\;\Pr\!\left(y_n=j\mid\bm{x}_n,\mathcal H_{<t_n}\right)
=\frac{\exp\eta_{nj}}{\sum_{j'\in\mathcal C(o_n)}\exp\eta_{nj'}},
\label{eq:softmax}
\end{equation}
with linear predictor
\begin{align}
\eta_{nj}={}& \underbrace{a_{s(j)}+b_{o_n,s(j)}}_{\text{attractiveness}}
+\underbrace{\bm{\gamma}_{h_n}^{\top}\bm{B}\!\big(D(o_n,j)\big)}_{
\text{in-vehicle distance profile}}
+\underbrace{\lambda\log\bar A^{(<t_n)}_{s(j),h_n,\tau_n}
+\lambda_B\log\bar B^{(<t_n)}_{s(j),h_n,\tau_n}}_{
\text{aggregate attractors}}
\nonumber\\
&+\underbrace{\mathbb{1}\{j\in\mathcal B\}\big[\delta_1
+\delta_2\,k/L_{\ell,d}
+\kappa\, e^{-r(o_n,j)/r_0}\big]}_{\text{through-terminal and
road-crossing utility}}
+\underbrace{\psi_1^{(w_n)}\tilde D_{nj}
+\psi_2^{(w_n)}\,j/L
+\psi_3\,\tilde T_n\tilde D_{nj}}_{\text{weather/temperature
interactions}}.
\label{eq:eta}
\end{align}
Each block has a direct interpretation. Station attractiveness
$a_s\sim\mathcal N(0,\sigma_a^2)$ is shared across all lines serving $s$,
positive values marking stations that draw alightings beyond what the
other covariates explain;
the origin-specific adjustment $b_{o,s}\sim\mathcal N(0,\sigma_b^2)$
individuates data-rich origins (median 102 chained observations) while
shrinking to zero for sparse ones (38\% of origins have fewer than 50), so
sparse origins inherit the system-wide pattern. The in-vehicle distance
profile uses a cubic B-spline basis $\bm B(\cdot)$ in kilometres with
five interior knots at equally spaced quantiles (5th--95th) of the
training-set distances, the basis shared between training and test; its
hour-bin--specific coefficients are partially pooled as
$\bm\gamma_h=\bar{\bm\gamma}+\bm\Delta_h$,
$\bm\Delta_h\sim\mathcal N(\bm 0,\sigma_\gamma^2\bm I)$, with
$\bar{\bm\gamma}$ the common distance profile and $\bm\Delta_h$ the
hour-bin deviation, reflecting the finding that trip-length distributions
shift with time of day without hard stratification. Positive $\lambda$
(resp.\ $\lambda_B$) raises the odds of stations that historically
receive many alightings (resp.\ boardings) at that hour-bin and day type.
The opposite-direction block combines a baseline offset $\delta_1$
(negative values penalize opposite-direction travel absent a crossing
motive), a boarding-position term $\delta_2\,k/L_{\ell,d}$ with
$k/L_{\ell,d}$ the relative position of the boarding stop along its
route, and the
displacement decay $\kappa\,e^{-r/r_0}$, whose bonus is maximal at
$r=0$ and vanishes as $r$ grows, so only candidates near the boarding
point receive it. Remaining priors follow the implementation:
$\sigma_a\sim\mathrm{Half}\text{-}\mathcal N(0,1)$,
$\sigma_b\sim\mathrm{Half}\text{-}\mathcal N(0,0.5)$, coordinates of
$\bar{\bm\gamma}\sim\mathcal N(0,2^2)$,
$\sigma_\gamma\sim\mathrm{Half}\text{-}\mathcal N(0,0.5)$,
$\lambda\sim\mathcal N(1,1^2)$, $\lambda_B\sim\mathcal N(0,1^2)$,
$\delta_1,\delta_2,\kappa\sim\mathcal N(0,2^2)$, and
$r_0\sim\mathrm{LogNormal}(\log 200\,\mathrm m,\,0.5^2)$.
Because any covariate constant across candidates cancels in
\autoref{eq:softmax}, weather and temperature cannot enter as main effects;
they appear only through candidate-varying interactions with the
standardized distance $\tilde D_{nj}=\big(D(o_n,j)/1000-4\big)/3$,
standardized temperature $\tilde T_n=(T_n-20)/5$, and route position
$j/L$, where $L$ is the stop count of the candidate's own direction
($L_{\ell,d}$ for $\mathcal A$, $L_{\ell,1-d}$ for $\mathcal B$), under
tight shrinkage
$\psi_1^{(w)},\psi_2^{(w)},\psi_3\sim\mathcal N(0,\sigma_\psi^2)$,
$\sigma_\psi\sim\mathrm{Half}\text{-}\mathcal N(0,0.1)$: the measured
effects are real but small ($+0.08$ stops in light rain), and
unregularized weather stratification degrades out-of-time performance in
one month of data.

For card-level personalization, let $c(n)$ denote the card making trip
$n$ and $n^{(<t)}_{c,o,j}$ denote card $c$'s \emph{posterior expected} prior choices of candidate $j$ from origin $o$, i.e., soft counts accumulated from the posterior destination distributions of the card's earlier trips (never from raw chain labels). A conjugate Dirichlet--multinomial layer, applied as a
post-estimation empirical-Bayes stage on top of the population posterior,
mixes the individual and population models:
\begin{equation}
q_n(j)\equiv\Pr\!\left(y_n=j\mid c(n)\right)
=\frac{n^{(<t_n)}_{c,o_n,j}+\tau_{\!c}\,p_n(j)}
{n^{(<t_n)}_{c,o_n,\cdot}+\tau_{\!c}},
\label{eq:card}
\end{equation}
where $n^{(<t_n)}_{c,o_n,\cdot}=\sum_j n^{(<t_n)}_{c,o_n,j}$, with
concentration $\tau_{\!c}>0$ selected by predictive
log-likelihood on the nested training split (a population model pre-fit on
the earlier training days scores candidate values on the later training
days it never saw); $\tau_{\!c}$ is therefore a validated
hyperparameter of the personalization stage rather than a jointly
estimated model parameter; it acts as an effective prior sample size, the
population model contributing $\tau_{\!c}$ pseudo-trips against the
card's accumulated soft counts. A card with no history reduces
exactly to the population model $p_n(j)$; a card with strong history is
pulled toward its own past behavior.

For chained trips, the recorded destination $z_n$ enters through a single
shared reliability model that deliberately does not depend on chain type,
chain rule, or matching distance (none of which exist at tap time):
\begin{equation}
\Pr\!\left(z_n\mid y_n=j\right)=\pi\,\mathbb{1}\{z_n=j\}
+(1-\pi)\,\big|\mathcal C(o_n)\big|^{-1},
\label{eq:noisy}
\end{equation}
i.e., with probability $\pi$ the chain identifies the true stop and
otherwise carries no information within the candidate set. Because $y_n$
ranges over a small finite set, it is marginalized analytically,
\begin{equation}
\Pr(z_n\mid\cdot)=\sum_{j\in\mathcal C(o_n)}
\Pr(y_n=j\mid\cdot)\,\Pr(z_n\mid y_n=j)
=\pi\,\Pr(y_n=z_n\mid\cdot)+\frac{1-\pi}{|\mathcal C(o_n)|},
\label{eq:marg}
\end{equation}
leaving a smooth likelihood with no discrete latent variables. Unchained
trips (19.6\%) contribute no $z$-likelihood; their destinations are
described entirely by the posterior of
\autoref{eq:softmax}--\autoref{eq:card}, so they enter the OD matrices with
appropriately wider uncertainty instead of being discarded.
\emph{Identifiability.} Without any ground-truth alightings, $\pi$ trades
off against the sharpness of the choice model and is not identified from
these data alone. We therefore fix $\pi$ on the grid
$\{0.50,0.70,0.90,1.00\}$ and report the sensitivity of predictive
performance, calibration, and the OD matrices; the endpoint $\pi=1$
recovers the conventional labels-as-truth approach and thus doubles as a
baseline. A secondary analysis places $\pi\sim\mathrm{Beta}(16,4)$ centered
on exact-stop chaining accuracies reported where entry--exit ground truth
exists \citep{assemi_improving_2020}, claiming that the data update this
prior weakly.

\subsubsection{Estimation and Posterior Inference}
\noindent The boarding and alighting NB models have separate parameter
blocks, indexed by $g\in\{B,A\}$,
$\bm\vartheta^{g}
={}\big(\beta_0^{g},\{u_s^{g}\},\{\rho_h^{g}\},\{\zeta_\tau^{g}\},
\{\xi_{s,h}^{g}\},\{\varepsilon_t^{g}\},\log\phi_0^{g},
\{\log\phi_s^{g}\},\sigma_u^{g},\sigma_\xi^{g},
\sigma_\varepsilon^{g},\sigma_\phi^{g}\big)$,
$\bm\psi=\big(\{\psi_1^{(w)}\},\{\psi_2^{(w)}\},\psi_3\big)$,
$\bm\theta
={}\big(\{a_s\},\{b_{o,s}\},\bar{\bm\gamma},\{\bm\Delta_h\},
\lambda,\lambda_B,\delta_1,\delta_2,\kappa,r_0,\bm\psi,
\sigma_a,\sigma_b,\sigma_\gamma,\sigma_\psi\big)$,
where $\bm\theta$ contains the choice-model parameters conditional on fixed
attractors and $\pi$.
Both NB models are fitted by Stochastic Variational Inference (SVI) to
complete station--hour--day panels, including zero counts. Their normalized
posterior-mean intensities form $m^{\mathrm{NB},B}$ and
$m^{\mathrm{NB},A}$ in \autoref{eq:attractors}; the concentrations
$\alpha_B$ and $\alpha_A$ are selected on a nested temporal validation
split. For each fixed
$\pi\in\{0.50,0.70,0.90,1.00\}$, the choice model is fitted to the
512{,}862 chained training trips using the marginalized likelihood
\autoref{eq:marg}. Minibatch SVI
uses mean-field factors for the high-dimensional station and
origin--destination effects and a full-rank Gaussian for the global
parameters. Reduced NUTS fits of the NB and choice models serve only as
diagnostics.

The alighting attractor and choice model are estimated through two
empirical-Bayes iterations, identical for every model variant and every
$\pi$ and starting from the same base attractors, so ablation and
sensitivity comparisons are controlled. Starting from chain-derived
alighting counts, each iteration fits $\bm\theta$, computes
\(
p_n^{z}(j)\propto p_n(j)\Pr(z_n\mid y_n=j),
\)
and refreshes $\bar A^{(<t)}$ using posterior expected counts, with its NB
profile refitted to posterior-sampled destinations. The boarding attractor
remains fixed because it uses observed boardings only. Thus, uncertainty
in $\bm\theta$ is conditional on the final attractors. Using a population
model pre-fit on the earlier training days, $\tau_{\!c}$ is selected on
the nested validation split those days never saw, and
card histories are constructed from earlier chained trips using their
$p_n^{z}$ soft counts. Hence, $\alpha_A$, $\alpha_B$, and $\tau_{\!c}$ are
selected hyperparameters of the empirical-Bayes stage, on which reported
uncertainty is conditional. In the
secondary analysis with $\pi\sim\mathrm{Beta}(16,4)$, the posterior is
augmented by $\pi$.

All historical features use only information available before prediction:
training features use days before $t_n$, test features are frozen at the
end of the training window, the outage day is excluded throughout, and
chain outputs, chain metadata, and same-day totals never appear as
prediction-time inputs. At
tap time, prediction uses the personalized distribution $q_n(j)$. For
retrospective OD estimation,
\(
\tilde p_n(j)\propto q_n(j)\Pr(z_n\mid y_n=j)
\)
when $z_n$ is available, and $\tilde p_n(j)=q_n(j)$ otherwise. Destinations
are sampled from $\tilde p_n$ for 100 posterior draws and aggregated into
OD flows $Q_{odt}$; their variation gives conditional posterior predictive
intervals, while differences across $\pi$ form reliability-scenario bands.
Agreement log-loss against $z_n$ is used only to compare models at the same
$\pi$, not to select $\pi$ or infer accuracy against true destinations.

\section{Results}\label{sec:results}

\noindent All models were fitted on the training window (May 1--24; outage
day May 9 removed; 512{,}862 chained trips) and evaluated on the test set (May 25--31; 152{,}202 label-scored trips) under an identical
estimation procedure, nested hyperparameter selection
($\alpha_A=300$, $\alpha_B=100$; $\tau_{\!c}=0.5$ for $\pi\le0.7$ and
$1.0$ for $\pi\ge0.9$), and common starting attractors. Entry-only
systems provide no true alightings, so all scores are computed
against trip-chain labels, and cross-$\pi$ comparisons are interpreted
under this constraint. Unless noted, results refer to the primary configuration $\pi=0.9$.

\begin{table}[htbp]
\caption{Out-of-time agreement with trip-chain labels, May 25--31
($n=152{,}202$; $\pi=0.9$).}
\label{tab:performance}
\centering
\small
\begin{tabularx}{\linewidth}{@{}Xcccc@{}}
\toprule
 \textbf{Models} & \textbf{Log-loss}\textsuperscript{a} & \textbf{Top-1} & \textbf{Top-3} & \textbf{Top-5} \\
\midrule
B1: Deterministic Trip Chaining \textsuperscript{b}
    & --- & --- & --- & --- \\
B2: Historical Destination Frequencies by Origin and Hour
    & 2.346 & 0.302 & 0.550 & 0.687 \\
B3: B2 with Card-History Personalization
    & 1.904 & 0.507 & 0.682 & 0.769 \\
HBLD without Card-History Personalization (Population Model)
    & 2.325 & 0.294 & 0.542 & 0.682 \\
HBLD with Card-History Personalization (Full Model)
    & \textbf{1.729} & \textbf{0.512} & \textbf{0.715} & \textbf{0.798} \\
\midrule
\multicolumn{5}{@{}l}{\itshape HBLD Full Model split by test-trip
type}\\
\quad on trips whose card has prior same-origin history ($n=90{,}148$)
    & 1.206 & 0.682 & 0.862 & 0.905 \\
\quad on cold-start trips without prior same-origin history ($n=62{,}054$)
    & 2.488 & 0.265 & 0.501 & 0.642 \\
\bottomrule
\multicolumn{5}{@{}p{\linewidth}@{}}{\footnotesize\textsuperscript{b}\,B1
defines the scoring target and covers only chained trips (80.4\%), so it
cannot be scored here. \newline
\footnotesize\textsuperscript{a}\,Lower log-loss and higher top-$k$ are better.
}
\end{tabularx}
\end{table}

\subsection{Predictive Performance}
\noindent\autoref{tab:performance} scores four models on the identical
test trips, ordered by the information they use. B2 uses aggregate
history only: the smoothed destination frequencies of each boarding stop
and hour-bin. B3 adds memorized individual habit: probability 0.75 on the
card's modal past destination when same-origin history exists, 0.25 on
B2. The HBLD population model uses the structural components of
\autoref{eq:eta} without personalization, and the full model adds the
card layer of \autoref{eq:card}. Against B3, the strongest baseline, the
full model wins distributionally rather than at the single best guess:
log-loss falls from 1.904 to 1.729 at comparable top-1 accuracy (51.2\%
vs.\ 50.7\%), while top-3 rises from 68.2\% to 71.5\% and top-5 from
76.9\% to 79.8\%, and distributional accuracy is what OD aggregation
with uncertainty requires. The final two rows split the same full
model's test trips by whether the card has prior same-origin history:
top-1 accuracy is 68.2\% with history and 26.5\% without, showing that
individual history dominates performance and motivating both the partial
pooling that governs its absence and the cold-start focus of the
ablation analysis that follows.

\subsection{Contribution of Aggregate Demand Information}

\noindent Two component-removal experiments assess the aggregate
attractors using
$\Delta=\mathrm{logloss}_{\mathrm{removed}}-
\mathrm{logloss}_{\mathrm{full}}$, where $\Delta>0$ favors the full model.
Intervals use paired trip-level differences with a day-clustered bootstrap
(\autoref{fig:results}a).

Removing the boarding attractor increases log-loss at every $\pi$, from
$0.0114$ nats per trip at $\pi=0.5$ $[0.0098,0.0125]$ to $0.0004$ at
$\pi=1.0$ $[0.0002,0.0008]$. Its contribution is larger when chain
evidence is discounted and when card history is unavailable: at
$\pi=0.9$, the increase is $0.0050$ for cold-start trips versus $0.0015$
for trips with history. Historical boarding demand therefore provides a
label-independent population signal that partly substitutes for individual
history.

Removing both attractors improves log-loss at $\pi=0.5$
($\Delta=-0.0051$ $[-0.0089,-0.0003]$), has no detectable effect at
$\pi=0.7$, and worsens it at $\pi=0.9$ ($0.0039$
$[0.0025,0.0060]$) and $\pi=1.0$ ($0.0047$). Because removing the boarding
attractor alone always harms performance, this reversal is associated with
the alighting attractor. Unlike the boarding attractor, which uses observed
counts, the alighting attractor is derived from chain evidence and treated
as fixed after its empirical-Bayes update. At low $\pi$, this plug-in
feature can retain more chain-derived structure than the likelihood is
intended to trust; at high $\pi$, the aggregate and trip-level evidence
instead reinforce each other.

The boarding coefficient is small and negative
($\hat\lambda_B=-0.009\pm0.005$, versus
$\hat\lambda=0.620\pm0.005$ for alighting), indicating a conditional
correction: stations generating many boardings are not necessarily strong
alighting destinations. Thus, label-free boarding information is the more
robust aggregate signal. Because the
evaluation target is itself trip-chained, these findings demonstrate
agreement and robustness across $\pi$, not accuracy against true
destinations.

\subsection{Sensitivity to Label Reliability}

\noindent Because evaluation uses chain outputs as proxy labels, agreement
metrics improve by construction as $\pi$ approaches one: log-loss decreases
from 1.945 to 1.712 and expected calibration error from 0.030 to 0.015
(\autoref{fig:results}b--c). These trends measure increasing agreement with
trip chaining, not accuracy against true destinations, and therefore cannot
be used to select $\pi$. Similarly, the apparent overconfidence relative to
chain labels at $\pi=0.5$ is expected because the model deliberately
discounts those labels.

Trip-level performance is largely insensitive once $\pi$ is high: top-1
accuracy differs by less than 0.05 percentage points between $\pi=0.9$ and
$1.0$. OD flows remain more sensitive because small probability shifts,
although insufficient to change the most likely destination, accumulate
across many trips. Among OD pairs with at least 50 expected trips, the
range of posterior-mean flow across $\pi$, relative to the pair's average
flow, is 13.3\% at the median and exceeds 27.1\% for the most sensitive
10\%. Thus, point predictions are stable at high $\pi$, but aggregate OD
estimates should retain the $\pi$-scenario bands.

Estimating $\pi$ does not resolve this uncertainty. Under
$\pi\sim\mathrm{Beta}(16,4)$, the posterior moves from a prior mean of 0.8
to 0.9956 (SD 0.0002). Without independently observed alightings, the large
sample identifies agreement with chain labels rather than their accuracy,
creating a precise but unvalidated estimate near one. External validation
is therefore required to identify $\pi$; absent such data, the scenario
analysis is the appropriate basis for OD reporting.


\subsection{Behavioral Parameters and Aggregate Agreement}

\noindent The estimated opposite-direction utility supports bus-assisted
road crossing. A positive proximity bonus
($\hat\kappa=3.93$) decays over
$\hat r_0=539\pm18$\,meters, substantially above its 200\,meters prior median, while
the baseline opposite-direction effect is negative
($\hat\delta_1=-3.51$). Thus, opposite-direction destinations are
penalized, but stops near the boarding location receive sufficient utility
to represent alighting on the opposite side, including stops beyond the
exact mirror. A NUTS diagnostic on the busiest line recovers the same
negative baseline and positive proximity effect, with
$r_0\approx473$\,m; other magnitudes are not directly comparable because
the diagnostic uses a line-specific subset.

\begin{figure}[htbp]
\centering
\includegraphics[width=\textwidth]{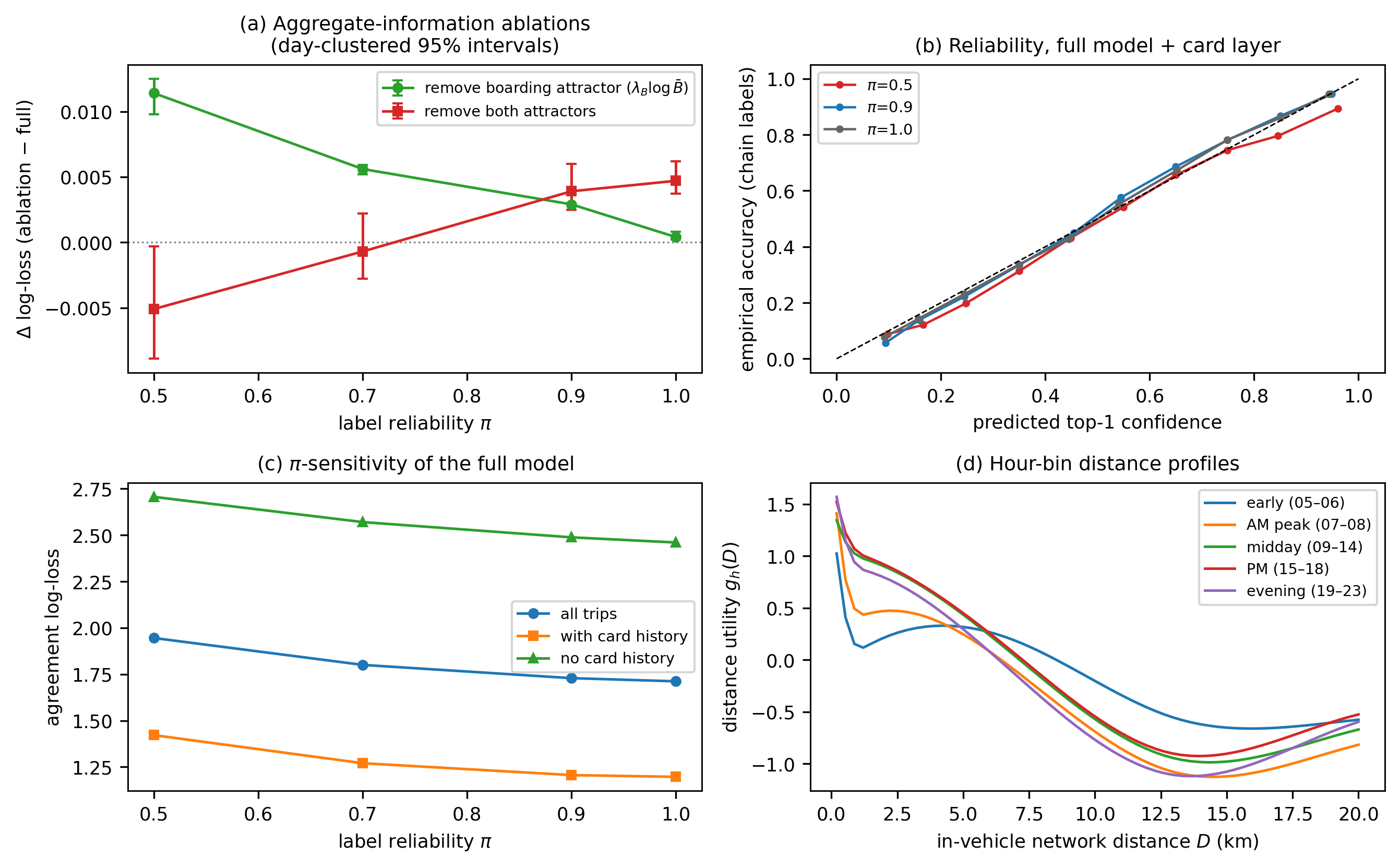}
\caption{(a) Paired attractor ablations across the reliability grid
(positive values indicate the removed term helps; day-clustered 95\%
intervals). (b) Reliability diagrams against chain labels. (c) Agreement
log-loss across $\pi$ by card-history stratum. (d) Posterior-mean
hour-bin distance-utility profiles.}\label{fig:results}
\end{figure}

The hour-specific distance profiles (\autoref{fig:results}d) show the
weakest distance deterrence before 07:00, consistent with the longer
early-morning trips observed in the exploratory analysis. At the aggregate
level, posterior station totals for chained test trips closely agree with
chain evidence (log-correlation 0.984); equality of their system-wide
totals is guaranteed by probability conservation. 

Weather coefficients require caution. Their near-identical values across
categories, together with $\hat\sigma_\psi=0.72$ despite a 0.1 prior scale,
indicate that the block absorbed an omitted route-position main effect.
Between-category deviations are small ($\lesssim0.25$), and the
temperature--distance interaction is weak
($0.001\pm0.006$). Thus, these coefficients are not interpreted as
weather effects.



\section{Discussion}\label{sec:discussion}
\noindent The findings show how aggregate demand and card history improve destination inference, while revealing how uncertain
trip-chain labels affect model evaluation and OD-flow uncertainty.

\subsection{Observed Boarding vs. Inferred Alighting Information}

\noindent Historical boarding demand improves prediction across all tested values of
$\pi$, with its largest contribution for trips lacking card history. It
therefore provides a robust population signal precisely where
personalization is unavailable. Alighting aggregates behave differently:
because they are reconstructed from chain outputs, their contribution
depends on how strongly those outputs are trusted. Although posterior soft
counts partially discount chain uncertainty, the resulting attractor is
treated as fixed in the next choice-model fit, so its uncertainty is not
fully propagated. More generally, label-derived aggregates can reintroduce
proxy-label structure through a plug-in channel even when the trip-level
likelihood discounts that structure. Directly observed boarding aggregates
avoid this dependence and are also less costly to construct.

\subsection{Population vs. Personalization}
\noindent The full HBLD model improves log-loss and top-$k$ agreement over the
card-history baseline while producing similar top-1 agreement. Its main
advantage is therefore a better destination distribution rather than a
different modal prediction, which is important because OD estimation
aggregates probability across all candidate destinations. Card history
remains the strongest trip-level signal, but partial pooling provides a
population fallback when it is unavailable. The remaining performance gap
for cold-start trips shows that partial pooling manages, rather than
eliminates, the cold-start problem and explains the particular value of
label-free boarding information for this group.

\subsection{Reliability vs. Uncertainty}
\noindent Because the evaluation target is the chain output, label-based metrics are
expected to favor $\pi=1$ and cannot identify accuracy against true
destinations. Likewise, the posterior estimate
$\pi\approx0.996$ measures consistency with the proxy labels rather than
their reliability; studies with observed entry and exit report exact-stop
rates nearer 72--80\% \citep{assemi_improving_2020}. Trip-level
performance is nearly unchanged between the tested settings
$\pi=0.9$ and $1.0$, but small probability shifts accumulate into
meaningful differences in OD flows. OD estimates should therefore be
reported with both conditional posterior predictive intervals and
$\pi$-scenario bands. The former remain conditional on the
empirical-Bayes attractors, card soft counts, and selected
$\alpha_A$, $\alpha_B$, and $\tau_{\!c}$, and consequently do not represent
total model uncertainty. Independent entry--exit or on-board observations
are required to identify $\pi$.

\section{Conclusion}\label{sec:conclusion}
\noindent This study developed a Hierarchical Bayesian Latent-Destination
(HBLD) model for inferring bus-trip destinations from tap-in-only smart-card
data. The model combines network structure, temporal patterns, historical
boarding and inferred alighting demand, and card-specific travel history.
Unlike conventional supervised approaches, it treats trip-chain outputs as
noisy evidence for a latent destination rather than as ground truth. It
therefore produces destination distributions for both chained and
unchained trips and propagates trip-level uncertainty into aggregate OD
flows.

Out-of-time evaluation shows that the full HBLD specification, combining historical boarding and alighting demands and personal card-history, improves agreement log-loss and
top-$k$ performance significantly. 
Card history is
the strongest individual signal, but observed boarding aggregates improve
prediction at every tested reliability level and contribute most to
cold-start trips, whose cards have no prior same-origin history. Inferred alighting aggregates are less robust because
their value depends on the assumed reliability of the same chain evidence
from which they are constructed. The model also identifies systematic
opposite-direction behavior consistent with through-terminal riding and
bus-assisted road crossing. Finally, trip-level agreement with chain labels changed little between the tested high-reliability settings ($\pi=0.9$ and $1.0$), although this does not establish accuracy against true destinations due to the absence of observed alightings.

These findings should be interpreted in light of several limitations.
Most importantly, no true alighting observations are available, preventing
direct validation of inferred destinations, evaluation of unchained trips,
or identification of the chain-reliability parameter $\pi$. The model also
relies on simplifying assumptions: destinations are restricted to
same-line downstream stops or opposite-direction stops reached through the
terminal; chain reliability is represented by one common $\pi$, with
errors distributed uniformly over the candidate set; and card behavior and
historical station activity are assumed sufficiently stable for past
patterns to inform later trips. Moreover, the empirical-Bayes attractors
are treated as fixed during choice-model estimation, so their uncertainty
is not fully represented in the reported intervals. Although trip-chain
outputs enter the model as noisy evidence, log-loss and top-$k$ metrics are
still evaluated against those same proxy labels. Their improvement as
$\pi$ increases therefore measures agreement with trip chaining, not
accuracy against true destinations. Finally, the one-month dataset from a
single city cannot establish geographic transferability or seasonality.
In particular, the same-weekday structure rejected with only sixteen
usable weekdays may become informative in longer panels.

Future work should first obtain entry--exit or on-board observations to
validate destinations, evaluate unchained trips, and identify $\pi$ without
chain-based scores. The model could then relax its assumptions through
context-specific reliability, non-uniform chain errors, flexible candidate
sets, dynamic card and station effects, and joint propagation of aggregate
uncertainty. An architecturally comparable neural network model (neural destination model) could also be evaluated using identical inputs, candidate sets, and temporal splits to isolate the value of Bayesian uncertainty and partial pooling.
Finally, longer multi-city and multimodal panels could test seasonality and
transferability while extending HBLD from route-level bus inference to
system-wide OD estimation across bus, subway, shared bicycle, and
intermodal trips.

\section{Acknowledgments}

\noindent The authors gratefully acknowledge Changzhou Public Transport for providing the original raw data used in this study. The authors also thank Nanjing Tianyu Technology Co., Ltd., and in particular Jie Li, for offering the source data and for providing helpful technical explanations that supported the authors’ understanding of the operational characteristics of the Changzhou bus system. The authors acknowledge the use of OpenAI ChatGPT, GPT-5.5 Thinking, to assist with language editing and LaTeX formatting. The AI tool was used only as a writing and formatting aid. All AI-assisted content was reviewed, revised, and verified by the authors. The authors take full responsibility for the study design, data processing, results analysis, conclusions, and final manuscript.


\section*{AUTHOR CONTRIBUTIONS}
\noindent The authors confirm contribution to the paper as follows: study conception and design: J. Ling, G. Zhao, Y. Su;
data collection: G. Zhao, J. Ling; analysis and interpretation of results: J. Ling, G. Zhao, Y. Su; draft manuscript preparation: J. Ling, G. Zhao, Y. Su. All authors reviewed the results and approved the final version of the manuscript.


\section*{DECLARATION OF CONFLICTING INTERESTS}

\noindent The authors declared no potential conflicts of interest with respect to the research, authorship, and/or publication of this article.




\section*{FUNDING}



\noindent The authors disclosed receipt of the following financial support 
for the research, authorship, and/or publication of this article: 
This research was supported by \emph{Independent Research Project of the State Key Laboratory of Intelligent Green Vehicle and Mobility, Tsinghua University} (grant no.~\emph{ZZ-GG-20250406}).



\clearpage

\nolinenumbers

\pagestyle{empty}
\thispagestyle{empty}

\renewcommand{\bibsection}{%
  \section*{REFERENCES}
  \vspace{12pt}
}

\setlength{\bibhang}{0.5in}
\setlength{\bibsep}{12pt plus 0.3ex}

\bibliographystyle{trb}
\bibliography{trb_template}

\end{document}